\documentclass[twocolumn, superscriptaddress, aps, 10pt]{revtex4-2}
\newcommand{\rev}[1]{{\color{black}#1}}
\usepackage{multirow}
\usepackage{array}
\usepackage{hyperref}
\usepackage{xcolor}
\usepackage{soul}
\usepackage{amsmath,amsfonts,amssymb}
\usepackage{mathrsfs}
\usepackage{verbatim}
\usepackage{graphicx}
\usepackage{outlines}
\usepackage{listings}
\usepackage{soul}
\begin{document}

\title{Resolving the Body-Order Paradox of Machine Learning Interatomic Potentials}
\author{Sanggyu Chong}
\email{sanggyu.chong@epfl.ch}
\affiliation{Laboratory of Computational Science and Modeling, Institute of Materials, \'Ecole Polytechnique F\'ed\'erale de Lausanne, 1015 Lausanne, Switzerland}
\author{Tong Jiang}
\affiliation{Department of Chemistry and Chemical Biology, Harvard University, Cambridge, Massachusetts 02138, United States}
\author{Michelangelo Domina}
\affiliation{Laboratory of Computational Science and Modeling, Institute of Materials, \'Ecole Polytechnique F\'ed\'erale de Lausanne, 1015 Lausanne, Switzerland}
\author{Filippo Bigi}
\affiliation{Laboratory of Computational Science and Modeling, Institute of Materials, \'Ecole Polytechnique F\'ed\'erale de Lausanne, 1015 Lausanne, Switzerland}
\author{Federico Grasselli}
\affiliation{Laboratory of Computational Science and Modeling, Institute of Materials, \'Ecole Polytechnique F\'ed\'erale de Lausanne, 1015 Lausanne, Switzerland}
\affiliation{(Present address): Dipartimento di Scienze Fisiche, Informatiche e Matematiche, Università degli Studi
di Modena e Reggio Emilia, 41125 Modena, Italy}
\author{Joonho Lee}
\affiliation{Department of Chemistry and Chemical Biology, Harvard University, Cambridge, Massachusetts 02138, United States}
\author{Michele Ceriotti}
\affiliation{Laboratory of Computational Science and Modeling, Institute of Materials, \'Ecole Polytechnique F\'ed\'erale de Lausanne, 1015 Lausanne, Switzerland}

\newcommand{\FB}[1]{{\color{teal} #1}}
\newcommand{\FBcancel}[1]{{\color{teal} \st{#1}}}

\newcommand{\SC}[1]{{\color{orange} #1}}
\newcommand{\TJ}[1]{{\color{blue} #1}}
\newcommand{\MD}[1]{{\color{green} #1}}

\begin{abstract}
In many cases, the predictions of machine learning interatomic potentials (MLIPs) can be interpreted as a sum of body-ordered contributions, which is explicit when the model is directly built on neighbor density correlation descriptors, and implicit when the model captures the correlations through non-linear functions of low body-order terms. In both cases, the ``effective body-orderedness'' of MLIPs remains largely unexplained: how do the models decompose the total energy into body-ordered contributions, and how does their body-orderedness affect the accuracy and learning behavior? In answering these questions, we first discuss the complexities in imposing the many-body expansion on \textit{ab initio} calculations at the atomic limit. Next, we train a curated set of MLIPs on datasets of hydrogen clusters and reveal the inherent tendency of the ML models to deduce their own, effective body-order trends, which are dependent on the model type and dataset makeup. Finally, we present different trends in the convergence of the body-orders and generalizability of the models, providing useful insights for the development of future MLIPs.

\end{abstract}

\maketitle

\section{Introduction}

The many-body expansion (MBE) expresses the global observable of a chemical system as a sum of contributions from interactions at different body-orders, where the ``bodies'' are taken to be atoms, molecules, or larger fragments in the system. The MBE enables the interpretation of complex interactions in terms of simpler body-ordered contributions, and can provide reasonable approximations to the global quantity, especially when the contributions decay rapidly with the number of bodies involved. The MBE has given rise to many local or fragment-based quantum chemistry methods \cite{Yang1991,Saebo1993,Yang1995,Baer1997,Kitaura1999,Dahlke2007,Solovyev2008,Richard2012,Pruitt2014} that offer favorable scaling for large systems, as well as force fields that can be used for atomic scale simulations at large length and time scales \cite{Rappe1992,Wang2004,Vanomm2010}.

Machine learning interatomic potentials (MLIPs) \cite{Behler2007,Bartok2010,Shapeev2016,drau19prb,Behler2021} enable \textit{ab initio}-quality atomistic simulations with linear system size scaling and low prefactors, further extending the accessible length and time scales of the simulations. MLIPs commonly adopt a locality ansatz based on the nearsightedness principle \cite{Kohn1996,Prodan2005}, and represent the local environments of chemical systems in terms of body-ordered correlations between the central atom and its neighbors.~\cite{Willatt2018,musi+21cr,Glielmo2021} This allows MLIPs to effectively capture the complex many-body interactions while retaining favorable scalability and transferability.

Here, one could draw a parallel between the MBE interpretation of physical observables and the correlation-based atomic representations of common MLIPs. Within this parallel, a paradox also emerges: how do the MLIPs make accurate predictions with a limited set of atomic body-order correlations, despite the MBE being exact only in the limit of all body-ordered contributions in the system (with their count approaching infinity in bulk systems)? This highlights a deeper gap in our understanding of the ``body-orderedness'' of MLIPs and its implications on their performance and learning behavior, especially for recent graph neural network (NN)-based models where the body-ordered correlations are implicit and their contributions are non-separable.

In this study, we systematically analyze the body-orderedness of NN-based MLIPs and the resulting learning behavior. First, we revisit the trends in the MBE of \textit{ab initio} calculations, which commonly serve as reference for MLIP training, at the atomic limit.
Next, we train a curated set of MLIPs on datasets of hydrogen clusters to investigate how the models infer body-order trends of hydrogen \cite{Cheng2020,Bischoff2024}, and how they compare to the reference. We further explore how the trends vary between the different MLIPs and dataset makeup, and discuss how they may affect model accuracy and learning behavior. Finally, we consider the relationship between the body-ordered energetics and the out-of-distribution accuracy or generalizability of MLIPs.

\section{Theory and methods}

\subsection{Parallelism between the MBE and the body-orderedness of MLIPs}{\MD{\label{sec:MBEandBOMLIPs}}}

In applying the MBE at the atomic limit, the total energy $E_A
$ of system \textit{A} with $N$ atoms is expressed as:

\begin{equation}\label{eq:mbe}
\begin{gathered}
E_A (\mathbf{r}_1, \mathbf{r}_2, ..., \mathbf{r}_N) = \sum_i V^{(1)} +
\sum_{i<j} V^{(2)}(\mathbf{r}_i, \mathbf{r}_j) \\
+ \sum_{i<j<k} V^{(3)}(\mathbf{r}_i, \mathbf{r}_j, \mathbf{r}_k) + \cdots + V^{(N)} (\mathbf{r}_1, \mathbf{r}_2, ..., \mathbf{r}_N)
\end{gathered}
\end{equation}

\noindent for atoms $i, j, k... \in A$ and their coordinates $\mathbf{r}$. The summations run over the canonically complete set of lower body-ordered ``sub-clusters'' that can be identified within $A$.
%For simplicity, we assume $A$ is mono-elemental.
The $m$-th body-ordered energy contribution $V^{(m)}$ % ($\neq E^{(m)}$)
from a sub-cluster with $m$ atoms can be expressed in terms of its total energy $E^{(m)}$ minus all the lower body-order terms:

\begin{equation}\label{eq:mbe_sub}
\begin{gathered}
V^{(m)}(\mathbf{r}_1, \mathbf{r}_2, ..., \mathbf{r}_m) = E^{(m)}(\mathbf{r}_1, \mathbf{r}_2, ..., \mathbf{r}_m) \\ - \sum_{1 \le k<m} \; \sum_{i'<j'<...< k'}V^{(k)}(\mathbf{r}_{i'}, \mathbf{r}_{j'}, ..., \mathbf{r}_{k'}).
\end{gathered}
\end{equation}

\noindent In the second term, the inner sum is same as those in Eq.~\eqref{eq:mbe}, and the outer sum runs over all body-orders $k$ lower than $m$. Here, we interpret $V^{(1)}$ to be $E_1$, the energy of an isolated atom. For $m=3$:

\begin{equation}\label{eq:mbe_sub3}
\begin{gathered}
V^{(3)}(\mathbf{r}_1, \mathbf{r}_2, \mathbf{r}_3) = E^{(3)}(\mathbf{r}_1, \mathbf{r}_2, \mathbf{r}_3) \\ - 3E_1 - V^{(2)}(\mathbf{r}_1, \mathbf{r}_2) - V^{(2)}(\mathbf{r}_1, \mathbf{r}_3) - V^{(2)}(\mathbf{r}_2, \mathbf{r}_3)
\end{gathered}
\end{equation}

\noindent The number of canonically complete sub-clusters for a given $m$ is $N!/(m!(N-m)!)$, and is the largest for $m \approx N/2$. The body-ordered (negative) \textit{force} contribution $\partial V^{(m)}/ \partial \mathbf{r}_i$ can be further derived from Eq.~\eqref{eq:mbe_sub} as follows:

\begin{equation}\label{eq:mbe_f_sub}
\begin{gathered}
\frac{\partial V^{(m)}}{\partial \mathbf{r}_i} = \frac{\partial E^{(m)}}{\partial \mathbf{r}_i} \\ - \sum_{2 \le k<m} \; \sum_{i''<j''<...< k''} \frac{\partial V^{(k)}(\mathbf{r}_{i''}, \mathbf{r}_{j''}, ..., \mathbf{r}_{k''})}{\partial \mathbf{r}_i}
\end{gathered}
\end{equation}

\noindent $\partial V^{(m)}/ \partial \mathbf{r}_i$ is a local quantity for atom $i$, and the inner summation runs over the neighboring atoms of $i$.

The locality ansatz of MLIPs leads to the following expression for the predicted total energy $\tilde{E}_A$:

\begin{equation}\label{eq:mlip_locale}
\begin{gathered}
\tilde{E}_A = \sum_{i\in A} \varepsilon_i = \sum_{i\in A} \varepsilon_i (\{\mathbf{r}_{ij}\}_{j \in A_i})
\end{gathered}
\end{equation}

\noindent $\varepsilon_i$ is the ``local'' energy associated with atom $i$, and it is a function of the local geometry described via $\mathbf{r}_{ij} = \mathbf{r}_i - \mathbf{r}_j$ for all $j$ in $A_i$, the local environment of atom $i$. The MLIPs differ in the functional form used to predict $\varepsilon_i$, which usually involves combining multiple $\mathbf{r}_{ij}$ terms to describe higher body-order correlations.~\cite{musi+21cr,Nigam2022} Among the models, there exists a subset~\cite{Shapeev2016,drau19prb} where features that formally depend on a fixed number of neighbors are used as the polynomial basis for a linear expansion:

\begin{equation}\label{eq:mlip_linear}
\begin{gathered}
\varepsilon_i = \sum_{\nu} \boldsymbol{\phi}_{\nu}(\{\mathbf{r}_{ij}\}_{j \in A_i}) \cdot \mathbf{w}_\nu
\end{gathered}
\end{equation}

\noindent Here, $\nu$ is the ``correlation order'' involving multiple neighbor atoms $j$ (then, body order equals $\nu+1$). The expression tells us that the predicted energy can be explicitly decomposed into body-ordered contributions. Despite the apparent parallel with the MBE (Eq.~\eqref{eq:mbe}), previous works~\cite{Chong2025,Ho2024,Tan2025} have shown that there does not exist a 1-to-1 correspondence between the two because $\phi_\nu$ also contains terms associated with lower body-order correlations, unless they are explicitly elimintated~\cite{Ho2024} (see Appendix~\ref{app:lin_bo_models}).

In this study, we focus on the NN-based MLIPs, which have shown greater accuracies for much larger portions of the chemical space,~\cite{Batatia2024,Park2024,Mazitov2025,Rhodes2025} yet their body-orderedness has never been considered in detail. For the NN-based models, $\varepsilon_i$ often involves nonlinearities that make it challenging to derive an analytical connection to the MBE. 
One of such models is the Behler-Parrinello NN (BPNN)~\cite{Behler2007}, where the local descriptors of 2- and 3-body correlations are taken as inputs to a fully connected feed-forward network that predicts $\varepsilon_i$, which are then summed across the system to obtain the total energy. Here, we specifically consider SOAP-BPNN, which adopts the Smooth Overlap of Atomic Position (SOAP)~\cite{Bartok2013} as the local descriptor of choice. Starting from the 3-body SOAP descriptors, SOAP-BPNN \emph{implicitly} reaches higher body-orders through the multiple NN layers and the nonlinear activation functions. Since SOAP is an incomplete descriptor~\cite{Pozdn2020}, the resulting SOAP-BPNN descriptor also exhibits incompleteness at higher body-orders.

% \begin{equation}\label{eq:mlip_mpnn}
% \begin{gathered}
% \mathbf{x}_i = \mathsf{x}_i^{l_\text{max}}, \; \; \; \;  \mathsf{x}_i^{l+1} = \psi(\mathsf{x}_i^{l}, \sum_{j}\xi(\mathsf{x}_j^{l},\mathsf{e}_{ij}))
% \end{gathered}
% \end{equation}

% \noindent where the second expression is applied recursively via multiple layers in the model architecture. $\mathsf{x}_i^l$ is the node embedding of atom $i$ at layer $l$, $\mathsf{e}_{ij}$ is the edge feature that encodes $\textbf{r}_{ij}$, $\xi$ is the aggregate function, and $\psi$ is the update function. The ultimate body-order is dictated by the initial node embedding $\mathsf{x}_i^0$, edge featurization ($\mathsf{e}_{ij}=f(\textbf{r}_{ij})$), total number of layers $l_\text{final}$, and the aggregate and update functions.

More recent MLIPs incorporate message-passing NNs~\cite{Gilmer2017} and transformers~\cite{Vaswani2023} in their architectures. One example that we consider is MACE~\cite{bata+22nips}, an equivariant message-passing NN-based MLIP grounded in the atomic cluster expansion (ACE) formalism.~\cite{drau19prb} In MACE, each layer describes the body-order correlations of the given atomic environment up to $\nu=3$. Equivariant message passing effectively introduces a new one-particle basis, raising $\nu$ by 1.~\cite{Nigam2022} With two of such layers, MACE \emph{explicitly} achieves maximum body-order of 13~\cite{Kovacs2023}. Lastly, we consider the point-edge transformer (PET)~\cite{Pozdn2023}, a transformer-based model that does not enforce exact rotational symmetry. In PET, the \texttt{softmax} function used in the attention mechanism, as well as the activation functions in the feedforward blocks, \emph{implicitly} lead to a theoretically infinite body-order.

Even though there is no explicit decomposition of the predictions of these nonlinear NN-based MLIPs into body-ordered contributions, the connection is quite strong, with low-order correlations being used as inputs to internal operations that raise the body-order ultimately perceived by the model. It is therefore interesting to compute, \emph{empirically} through Eq.~\eqref{eq:mbe_sub} and~\eqref{eq:mbe_f_sub}, how the MBE of the trained models compare to that of the reference electronic structure methods, how it depends on the training details, and how it affects the transferability of the models.

\subsection{Hydrogen cluster sampling and calculation details}

In this study, we construct and utilize datasets of hydrogen octamers (8-mers) for MLIP training and analysis. Given the simple electronic structure of hydrogen, this ensures that the analysis can be focused on the body-order-specific effects as much as possible, and reduces the cost of calculations that probe the body-ordered energetics beyond the density functional theory (DFT) level (see Section \ref{sec:ref_bo}).

The 8-mers are sampled from the bulk hydrogen datasets of Cheng et al.~\cite{Cheng2020} We construct two distinct datasets: a ``high density'' (high $\rho$) dataset in which the clusters are sampled from 100 configurations with the highest density (average $\rho=$ 1.34 g/cm$^3$), and a ``low density'' (low $\rho$) dataset where the clusters are sampled from 100 configurations with the lowest density (average $\rho=$ 0.461 g/cm$^3$). In the high $\rho$ dataset, each 8-mer is sampled by choosing a hydrogen at random from a given configuration, then taking its seven nearest neighbors. In the low $\rho$ dataset, sampling is performed by taking a random hydrogen atom and its closest neighbor, then adding three nearest \textit{pairs} of atoms to complete the 8-mer. Example clusters are shown in Figure~\ref{fig:ref_bo}. Such difference in the sampling protocol ensures the sampling of distinct chemical trends seen in the corresponding bulk phases, where high $\rho$ systems are atomic, covalently bound and metallic, and low $\rho$ systems are molecular, bound by non-covalent interactions, and insulating. From each dataset, we randomly select 500 clusters, for which we also create the accompanying datasets containing canonically complete sets of their sub-clusters.

To obtain the energies and forces for MLIP training, DFT calculations are performed using FHI-aims~\cite{Blum2009}. The PBE exchange correlation functional~\cite{PBE} is employed with the ``tight'' species-default basis set, and Gaussian smearing with $\sigma=0.025$ eV is used to determine the occupations. 
To assess body-ordered energetics of hydrogen clusters at a much higher level of theory, we also perform spin-adapted density matrix renormalization group (DMRG) calculations~\cite{Sharma2012} using \textsc{Block2}~\cite{zhai2023block2} on a few representative clusters, using one- and two-electron integrals generated by \textsc{PySCF}~\cite{sun2018pyscf,sun2020recent}.
During method investigation, coupled-cluster with single, double, and triple excitations (CCSDT) calculations were also carried out in \textsc{Q-Chem}~\cite{epifanovsky2021software} for internal checks.
Further details of the cluster sampling protocol and the reference calculations are given in the Supplementary Material.

\subsection{MLIP training details}\label{subsec:mlip-training}

We consider three NN-based models widely varying in their architectures: SOAP-BPNN, MACE, and PET. In all cases, models are trained via stochastic gradient descent with the Adam optimizer~\cite{Kingma2017} on the mean squared loss of energies and forces. Train, validation, and test splits of the datasets are kept consistent between the models.
We set a consistent cutoff radius of 5.5 Å in all three models, which corresponds to the value used in the MLIPs developed in the previous work and encompasses the entire cluster.~\cite{Cheng2020} In SOAP-BPNN and MACE, the baseline energy, i.e. the energy of the isolated atom as perceived by the model, is fixed to the reference isolated atom energy. In PET, the existence of central tokens and their incorporation into the attention mechanism prevents the model from having a strict, pre-defined baseline. Instead, we take a data-driven approach and include an isolated atom reference configuration in the training sets for PET, allowing the model to learn an isolated atom energy to a good accuracy ($< 0.01$ eV from the reference value). We note that any deviation in the learned isolated atom energy from the reference may give PET a small advantage in the learning tasks. \rev{We have seen, however, that allowing the MLIPs to adjust freely their baseline energy does not impact significantly our observations (see Supplementary Material).} The ratio between the energy and force losses is fixed at 1:1 for SOAP-BPNN and PET. For MACE, we adopt its default training routine, which is to use 1:100 ratio between energy and force losses for the first stage, and 1000:100 in the second stage. The rest of the training details and hyperparameters are set to the defaults of the models. Full sets of the inputs and hyperparameters are provided in our Materials Cloud repository.~\cite{Talirz2020}

\section{Body-ordered energetics of hydrogen clusters}\label{sec:ref_bo}
\begin{figure}[t]
    \centering    \includegraphics[width=\columnwidth]{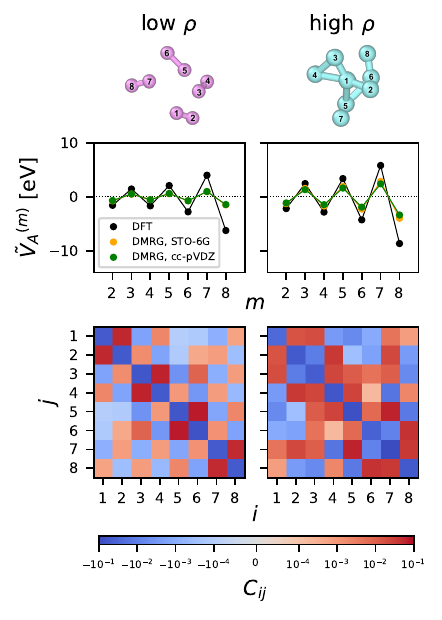}
    \caption{Body-ordered energetics and spin-spin correlations for a sample low $\rho$ 8-mer (left) and a sample high $\rho$ 8-mer (right). The atomic configurations of the sampled clusters are presented in the first row. Plots of the body-ordered energetics probed by $\tilde{V}_A^{(m)}$ of Eq.~\eqref{eq:norm_sub} are presented in the second row for the DFT and DMRG calculations. The spin-spin correlations results in the third row are obtained from the DMRG calculations, and are plotted in a symmetric log scale with a linearity threshold of $10^{-4}$. Atomic indices used in these plots are shown in the first row.}
    \label{fig:ref_bo}
\end{figure}

Before any MLIP analysis, we first evaluate the body-ordered energetics of hydrogen 8-mers in the reference \textit{ab initio} calculations. In many cases, the MBE formalism is applied at the \emph{molecular} level~\cite{Riera2020,Riera2020b,Bull2023}, where $V^{(m)}$ is expected to converge quickly to zero with increasing $m$. An archetypal system is water~\cite{Gora2011,Babin2012,Medders2013,Medders2013b,Richard2014,Pinski2014,Lao2016,Paesani2016,Heindel2021,Zhai2023}, for which many-body force fields have shown good accuracies with limited body-orders, up to 4, including also a polarizable model baseline. Even for water, however, body-ordered energetics at the DFT theory-level can converge slowly, and show large oscillations when ions are introduced.~\cite{Ouyang2014,Broderick2024} \rev{In the context of MLIPs and their body-orders, where body-ordered correlations or other geometric features are computed between atoms, the consistent scale at which to apply the MBE  formalism is also that of individual \emph{atoms}.}
At the atomic level, where covalent and metallic interactions dominate, even more complex and non-trivial trends have been previously observed for mercury, sodium, silicon, and gold.~\cite{Paulus2004,Hermann2007}

Here, we quantify the atomic MBE convergence for the hydrogen 8-mers used in this study. For all body-orders from 2 to 8, we compute:

\begin{equation}\label{eq:norm_sub}
\begin{gathered}
\tilde{V}^{(m)}_A = \frac{\sum_{i<j<...<m}V^{(m)}(\textbf{r}_i, \textbf{r}_j, ...,\textbf{r}_m)}{\frac{N!}{m!(N-m)!} \cdot m}
\end{gathered}
\end{equation}

\noindent For each $m$, energy contributions from the canonically complete set of $m$-mers are summed, then normalized by their total count and $m$ to yield the average body-ordered energy contribution, per atom. This allows for a ``fair'' comparison between the body-ordered contributions without any effects associated with the number of sub-clusters, which is largest when $m \approx N/2$. The resulting trends for sample 8-mers from low $\rho$ and high $\rho$ datasets are shown in Figure~\ref{fig:ref_bo}.

In both cases, an oscillatory and divergent trend is observed in the DFT body-ordered energetics (black markers). A negative energy contribution is first observed at $m=2$, then a larger positive contribution at $m=3$, and the sign of $\tilde{V}^{(m)}_A$ continues to alternate with the magnitude increasing with $m$. The anticipated chemical trend between the low $\rho$ and high $\rho$ 8-mers is manifested as a difference in the magnitudes of $\tilde{V}^{(m)}_A$ across all $m$, with larger contributions observed for the high $\rho$ 8-mers over the low $\rho$ 8-mers. Similar trends persist with the exact correlation treatment using DMRG, revealing that the apparent trend is not a mere consequence of the approximate nature of DFT~\cite{Broderick2024} (see Appendix~\ref{app:bo_frag} for further consideration of the DFT self-interaction error).
%, and originates from the highly non-local electronic correlations present in the distorted geometries, which arise from applying the MBE at the atomic level.

Previously, long-range many-body interactions have been observed in 1D hydrogen chains of up to 50 atoms~\cite{Motta2020}, where strong antiferromagnetic (AFM) spin-spin correlations are present. Here, we also analyze spin-spin correlations $C_{ij}$ for representative 8-mers in three spatial dimensions,
where $C_{ij} = \langle \hat{n}_{i\uparrow} \hat{n}_{j\downarrow}\rangle - \langle \hat{n}_{i\uparrow} \rangle \langle \hat{n}_{j\downarrow}\rangle$.
Figure~\ref{fig:ref_bo} reveals that both low and high $\rho$ 8-mers display substantial spin-spin correlations across all atomic pairs, thereby contributing to the apparent non-convergence of the body-ordered energetics.
The correlation patterns reveal distinct behaviors between the two density regimes: 
in the low $\rho$ 8-mer, strong correlations are present between bonded atom pairs, 
and non-negligible correlations also persist between the intermolecular pairs.
The high $\rho$ 8-mer exhibits stronger and more delocalized spin correlations across all pairs within the cluster.

These results demonstrate that the ``true'' body-ordered energetics of hydrogen 8-mers are inherently oscillatory and non-converging. We note that the contrast between our results and the mathematically proven exponential convergence of body-orders~\cite{Thomas2022} arises from the choice of $E_1$ in the MBE. While our expansion references itself to the ``vacuum'' (i.e., each sub-cluster is considered in complete isolation), considerations in Ref.~\cite{Thomas2022} assume full awareness of the \textit{entire} local environment even for the sub-clusters, \rev{which is more consistent with the stated goal of investigating the convergence of MLIPs, even though it leaves the definition of $E_1$ somewhat vague.} In Appendix~\ref{app:convergence}, we further unravel the dependence of the convergence trends on the baseline choice, and rationalize that for energy-stable atomic clusters such as the hydrogen 8-mers of our study, the observed oscillatory trend in the vacuum-referenced MBE is reasonable under the choice of $E_1$ as the isolated atom energy.

\begin{figure*}
    \centering
    \includegraphics[width=\textwidth]{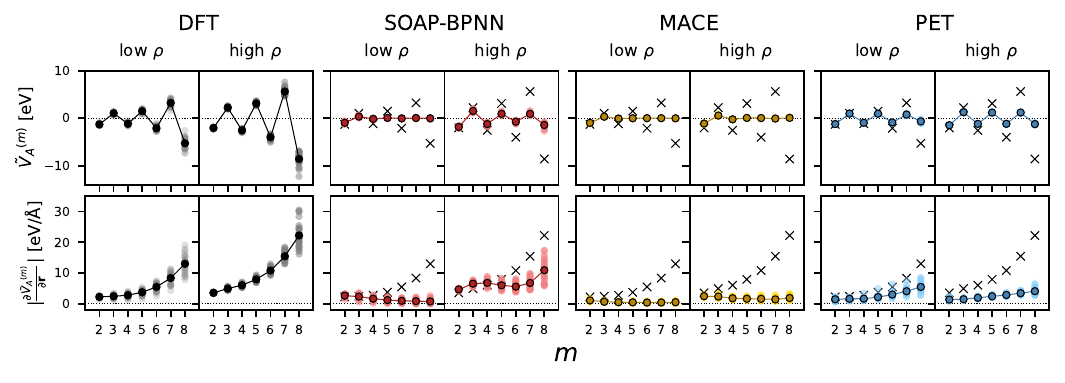}
    \caption{Body-order trends of 25 low $\rho$ and 25 high $\rho$  hydrogen 8-mers computed with DFT and MLIPs, as probed by $\tilde{V}^{(m)}_A$ of Eq.~\eqref{eq:norm_sub} and $| \partial \tilde{V}^{(m)}_A / \partial \textbf{r} |$ of Eq.~\eqref{eq:norm_sub_f}. The raw distribution of values are shown in lighter markers, and the mean values across the samples are shown with a darker outlined marker. The DFT mean values are presented as crosses in the plots for the MLIPs.}
    \label{fig:mlip_raw_bo}
\end{figure*}

In light of the strong dependence of the body-ordered energetics on the choice of the baseline energy $E_1$, it might be more appropriate to evaluate the convergence of the expansion by looking at the magnitude of the \emph{forces}, that are independent of the choice of $E_1$. 
We compute the following quantity
\begin{equation}\label{eq:norm_sub_f}
\begin{gathered}
\left|\frac{\partial \tilde{V}^{(m)}_A}{\partial \textbf{r}} \right| = \frac{ \sum_{A^{(m)} \in A} \sqrt{\frac{1}{m}\sum_i \left| \frac{\partial V^{(m)}}{\partial \mathbf{r}_i} \right|^2}}{\frac{N!}{m!(N-m)!}}
\end{gathered}
\end{equation}
The numerator is the sum of root mean square (RMS) of the norms of $\partial V^{(m)}/ \partial \mathbf{r}_i$ for the canonically complete set of $m$-mers, where the RMS of norms is computed over the atoms of a given $m$-mer. The denominator performs a normalization analogous to Eq.~\eqref{eq:norm_sub}.
As shown in the first panel of Fig.~\ref{fig:mlip_raw_bo}, for DFT the magnitude of the BO forces does not converge by $m=8$, indicating that the slow convergence is not only a consequence of the choice of a vacuum reference for the expansion, but reflects the high degree of electronic correlations for many of the sub-clusters.

\section{Effective body-orderedness of MLIPs}\label{sec:mlip_bo}

We now analyze the ``effective'' body-orderedness of MLIPs by training SOAP-BPNN, MACE, and PET on an hydrogen 8-mer dataset and computing $\tilde{V}^{(m)}_A$ and  $| \partial \tilde{V}^{(m)}_A / \partial \textbf{r} |$ from Eq.~\eqref{eq:norm_sub} and~\eqref{eq:norm_sub_f}. The dataset includes both low $\rho$ and high $\rho$ 8-mers in a 1:1 ratio. 10,000 8-mers split in 8:1:1 proportions are used as the training, validation, and test sets, with stratification between the $\rho$ regimes. All resulting models show energy and force RMSEs below 0.025 eV/atom (6.2 \%RMSE) and 0.375 eV/Å (15.6 \%RMSE, see Table S1). The analysis is performed on 25 low $\rho$ and 25 high $\rho$ 8-mers in the test set, for which the DFT reference values are available.

%We compute both the mean body-ordered energies~\eqref{eq:norm_sub} and the force magnitude~\eqref{eq:norm_sub_f}. {\color{red} comment also on the energy convergence. conclude saying well use only forces from now on}
Figure~\ref{fig:mlip_raw_bo} reveals that all MLIPs deviate from the DFT body-order trends in both energies and forces and infer their own, effective body-orders for the hydrogen 8-mers, with much smaller magnitudes across all $m$.
In SOAP-BPNN, body-orders of the low $\rho$ 8-mers are fast-converging with $m$, whereas those of the high $\rho$ 8-mers exhibit an oscillatory, slow-converging behavior in $\tilde{V}^{(m)}_A$ and an overall increasing trend in $| \partial \tilde{V}^{(m)}_A / \partial \textbf{r}|$. The spread of values across individual samples is also much narrower for low $\rho$ 8-mers compared to the high $\rho$ 8-mers that reach a standard deviation ($\sigma$) of 0.481 eV and 3.704 eV/Å in energies and forces for $m=8$.

In MACE, the body-order trends of $\tilde{V}^{(m)}_A$ is fast-converging for both densities with the significant contributions limited to $m \leq 4$, and $|\partial \tilde{V}^{(m)}_A / \partial \textbf{r}|$ also exhibits the lowest values across all $m$. The body-ordered contributions of both energies and forces are larger in the high $\rho$ 8-mers than the low $\rho$ 8-mers. The spread across individual samples is the narrowest among all MLIPs, with an average $\sigma$ of 0.060 eV and 0.434 eV/Å observed for high $\rho$ 8-mers. This suggests that the effective body-ordering of MACE is applicable across many samples with minimal variation. In PET, $\tilde{V}^{(m)}_A$ shows an oscillating, non-converging trend and $| \partial \tilde{V}^{(m)}_A / \partial \textbf{r} |$ increases with $m$ for both low and high $\rho$ 8-mers. While the magnitudes of $\tilde{V}^{(m)}_A$ are higher for the high $\rho$ 8-mers over the low $\rho$ 8-mers, those of $| \partial \tilde{V}^{(m)}_A / \partial \textbf{r} |$ exhibit the reverse trend, which is the opposite of the reference. PET shows moderate spread of values compared to the other two models, where the largest $\sigma$ values are 0.239 eV and 1.475 eV/Å are observed for low $\rho$ 8-mers.

\begin{figure*}
    \centering
    \includegraphics[width=\textwidth]{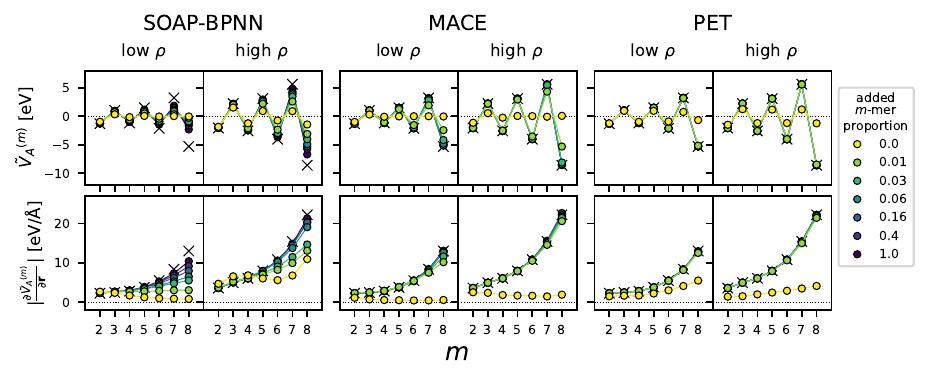}
    \caption{Changes in the body-order trends of MLIPs when explicit resolution to the DFT reference is attempted. Mean $\tilde{V}^{(m)}_A$ and $| \partial \tilde{V}^{(m)}_A / \partial \textbf{r} |$ values across 25 low $\rho$ and 25 high $\rho$ test set hydrogen 8-mers are shown. Multiple plots are made for different $m$-mer proportions, where the proportion corresponds to the number of $m$-mers added over all available $m$-mers for body-order resolution. In each panel, reference mean $| \partial \tilde{V}^{(m)}_A / \partial \textbf{r} |$ values from DFT are marked with crosses.}
    \label{fig:mlip_bo_resolve}
\end{figure*}

\begin{figure}[h!]
    \centering
    \includegraphics[width=\columnwidth]{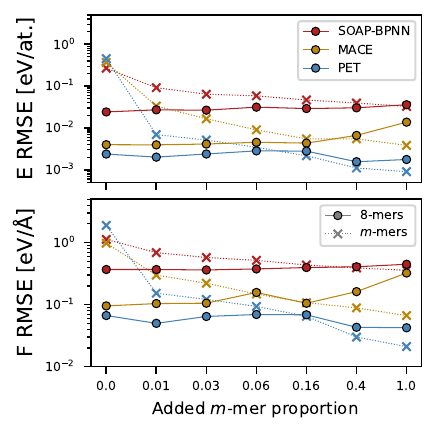}
    \caption{Energy and force RMSEs of the MLIPs computed for the full structure 8-mers (filled circles) and the sub-cluster $m$-mers (crosses) under different degrees of body-order resolution achieved with the addition of $m$-mers to the training set.}
    \label{fig:rmse_bo_resolve}
\end{figure}

In Figure S1, we present the ``per-$m$'' RMSEs of the three MLIPs on the \emph{sub-clusters} of the test set 8-mers, assessing the accuracy of the effective body-ordered energetics of the MLIPs. All three models exhibit similarly large RMSEs on average for both energies (0.325 eV/atom) and forces (1.21 eV/Å) for all $m<8$. This corroborates that the intuited body-ordered energetics are only effective, and significantly far from the DFT reference. \rev{In the Supplementary Material, we also explore the body-order trends of the MLIPs when $E_1$ is no longer fixed to the isolated atom energy. We learn that all MLIPs retain similar trends with notably smaller contributions across all body-orders, with the exception of PET exhibiting slightly larger $| \partial \tilde{V}^{(m)}_A / \partial \textbf{r} |$ values for both low $\rho$ and high $\rho$ 8-mers.}

\section{Explicit resolution of reference body-orders}\label{sec:bo_resolv_ref}

Next, we re-train the MLIPs on augmented datasets that aim to resolve their body-orders to the DFT reference. Canonically complete sets of $m$-mers, $2 \le m < 8$, for 200 low $\rho$ and 200 high low $\rho$ 8-mers (98,400 new structures in total), are added to the training set. The same is done for the validation and test sets, with 25 low $\rho$ and 25 high $\rho$ 8-mers each. Multiple models are trained at different augmentation proportions (inclusion of all 98,400 $m$-mers is 1), and in doing so, stratification is performed so that the complete set of canonical sub-clusters for a given 8-atom configuration are all included at once. 

\begin{figure*}
    \centering
    \includegraphics[width=\textwidth]{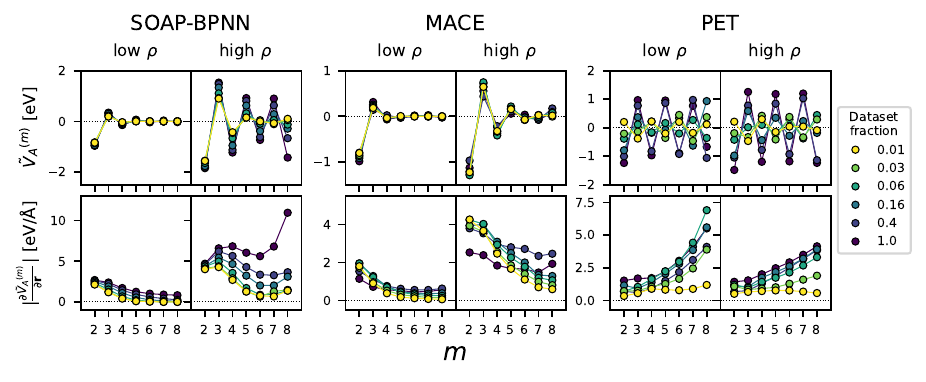}
    \caption{Body-order trends of the MLIPs trained on different fractions of the original training dataset. Mean $\tilde{V}^{(m)}_A$ and $| \partial \tilde{V}^{(m)}_A / \partial \textbf{r} |$ values over 25 low $\rho$ 8-mers and 25 high $\rho$ 8-mers are separately plotted. Note that the $y$-axis ranges are tailored for each model to focus on their individual learning behavior.}
    \label{fig:mlip_lc_bo}
\end{figure*}

Figure~\ref{fig:mlip_bo_resolve} shows that the DFT body-order trends are quickly captured by both MACE and PET as the $m$-mers \rev{are added} to the training set. In fact, for both $\rho$ 8-mers, near-complete resolution is achieved for $m \leq 6$ at 0.01 added $m$-mer proportion, and further increase in the proportion quickly resolves the $6 \leq m \leq 8$ contributions, faster for PET than MACE. This proves that while the MACE and PET architectures have sufficient flexibility to learn the reference body-order trends, without explicit resolution of the body-orders, their inherent tendencies result in trends that are drastically different from the reference. In SOAP-BPNN, $m \leq 4$ contributions are also resolved quickly at 0.01 added $m$-mer proportion, akin to the other two models. SOAP-BPNN, however, struggles to capture the higher body-order contributions at all added $m$-mer proportions, although it still gradually approaches the reference values. Better resolution is achieved for high $\rho$ 8-mers than the low $\rho$ 8-mers. In all MLIPs, the resolution takes place consistently faster for lower $m$, which must be related to how the dimensionality of the potential energy surfaces of $m$-mers exponentially increases with $m$.

Figure~\ref{fig:rmse_bo_resolve} shows the energy and force RMSEs for the 8-mers and their \rev{fragments}. The RMSEs for the $m$-mers decrease monotonically for all MLIPs as more and more sub-clusters are augmented to the dataset, as expected. For the 8-atom structures, SOAP-BPNN and MACE exhibits a compromise in the accuracy of the full configurations with addition of the sub-cluster $m$-mers in the training dataset. The RMSEs generally increase as more $m$-mers are incorporated, with MACE exhibiting a more pronounced increase---by a factor of 3.4 in both the energy and force RMSEs upon full augmentation. In PET, the RMSEs eventually decrease to values that are lower than those before any $m$-mers are introduced to the dataset. \rev{Among the considered MLIPs, PET is the only architecture for which learning the reference body-ordered energetics further improves the accuracy on the full structures.}

\section{Body-ordered interpretation of MLIP learning dynamics}

Having established that \rev{in the absence of explicit fragments in the training set,} the MLIPs infer their own body-order trends, we now investigate how these trends depend on the composition of the training dataset. We first train the models on the sub-sampled versions of the original training set from Sec.~\ref{sec:mlip_bo} in order to probe the learning dynamics of the MLIPs in the context of body-orders. We use dataset proportions ranging from 0.01 to 0.4, and keep the validation and test sets fixed. The learning curves are shared in Figure S3. We also repeat the exercise for datasets exclusively composed of either 10,000 low $\rho$ 8-mers or 10,000 high $\rho$ 8-mers. The resulting model accuracies for the latter two cases are shared in Table S2 and S3.

Figure~\ref{fig:mlip_lc_bo} presents the body-order trends for the MLIPs trained on different proportions of the original 8-mer dataset of Sec.~\ref{sec:mlip_bo}. In SOAP-BPNN, the model intuits an overall converging body-order trend at 0.01 dataset proportion for both density regimes, while assigning larger magnitudes for the higher $\rho$ 8-mers. As the dataset expands, the initial, converging trend is largely retained in the low $\rho$ 8-mers, with only slight increase in the $| \partial \tilde{V}^{(m)}_A / \partial \textbf{r} |$ values for the larger body-orders. For the high $\rho$ 8-mers, the model steadily increases the contributions for $3 \leq m \leq 8$ until reaching the final observed trend at the full dataset size. Note that the $m=2$ contribution remains ``pinned'' in both $\rho$ 8-mers.

MACE exhibits a fast-converging body-order trend at the smallest dataset proportion. In $\tilde{V}^{(m)}_A$, this initial profile is kept constant with minor fluctuations for both low $\rho$ and high $\rho$ 8-mers under all dataset proportions. Even in $| \partial \tilde{V}^{(m)}_A / \partial \textbf{r} |$, the initial trend is largely kept constant in the low $\rho$ 8-mers for all dataset sizes, and is retained up to 0.16 dataset proportion in the high $\rho$ 8-mers. Past this proportion, the high $\rho$ 8-mers start to exhibit relatively larger deviations in $| \partial \tilde{V}^{(m)}_A / \partial \textbf{r} |$ from the initial trend. Contrary to the other two MLIPs, PET does not display any converging trend, and exhibits non-negligible contributions across all $m$ even at the smallest dataset proportion. As the dataset is further expanded, PET freely adapts with no discernible trend in $\tilde{V}^{(m)}_A$, and quickly increases the contributions from higher $m$ in $| \partial \tilde{V}^{(m)}_A / \partial \textbf{r} |$, in stark contrast with the behavior of other two models.

\begin{figure*}[t!]
    \centering
    \includegraphics[width=\textwidth]{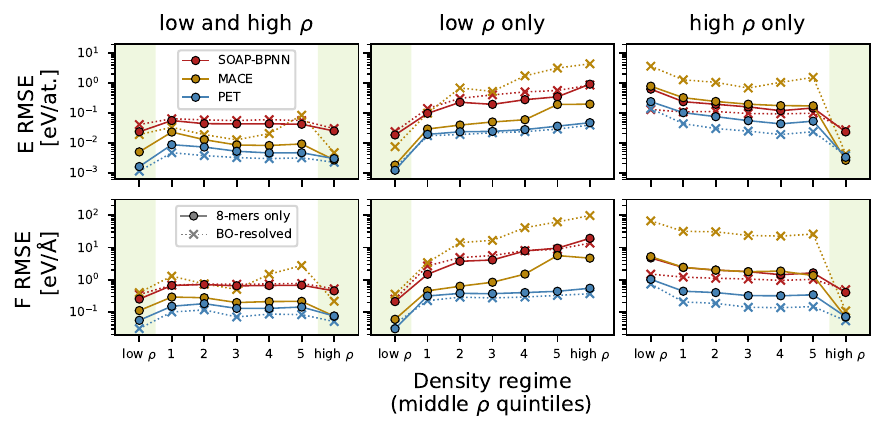}
    \caption{The energy (top) and force (bottom) RMSEs of the MLIPs computed across the entire density range. $x$ axis denotes the different density regimes for which the RMSEs have been computed. The two extrema are the original low $\rho$ and high $\rho$ regimes used in model training, and the five enumerated middle ticks correspond to the quintiles of the out-of-distribution intermediate density regimes. Each column corresponds to a different training dataset combination, which is shaded in light green. Circles show the RMSEs for the models trained on the 8-mers only, and the crosses show the RMSEs for the models trained on the 8-mers and their sub-clusters.}
    \label{fig:mlip_extrapol}
\end{figure*}

Most notably, we observe the tendency of MACE to prioritize the use of lower body-orders in its learning, which is also faintly present in SOAP-BPNN, and absent in PET. Similar behaviors persist in when the models are exclusively trained on the low $\rho$ 8-mers or the high $\rho$ 8-mers (see Figures S4-S7). MACE consistently prefers a fast-converging body-order trend, while SOAP-BPNN exhibits a converging trend for the low-density-only case and does not for the high-density-only case, and PET continues to show entirely arbitrary trends, especially in $\tilde{V}^{(m)}_A$. We attribute this distinct trend of MACE to the \textit{over-representation} of the lower body-orders (see Appendix \ref{app:mace_overrep}), which encourages the model to prioritize the use of lower body-ordered in formation in the learning task. We conjecture that the over-representation of lower body-orders restricts the use of the higher body-ordered features or descriptors to its full capacity, resulting in the limited capability of MACE to optimize its body-orders as the dataset is further expanded. 

\rev{One may wonder how much these trends depend on the details of an architecture and the choice of hyperparameters. As we show in the Supplementary Material, increasing the number of channels or correlation order $\nu$ of MACE modulates the body-order trend but doesn't change the overall fast-decaying tendency. Using nonlinear interaction blocks~\cite{Batatia2025} in MACE alleviates the model from converging trends and induces an oscillatory, diverging behavior in $\tilde{V}^{(m)}_A$ more similar to PET. Changing the token size for PET leads to changes in the body-order terms but not in their qualitative behavior, which remains oscillatory and without a clear relation to that of the underlying DFT reference.}

\section{Extrapolative performance of MLIPs}

One common interpretation of the MBE is that once sufficient convergence is reached in the body-ordered energetics, the expansion can be generalized to any applicable system with good accuracy. To assess this, we evaluate the extrapolative performance of the MLIPs, which have shown varying trends of convergence in their effective body-ordered energetics with respect to $m$. The energy and force RMSEs are computed for a dataset of out-of-distribution 8-mers that have been further sampled from the intermediate density regime of the Cheng et al.~\cite{Cheng2020} bulk hydrogen dataset, between the two density extrema from which the original 8-mers were sampled. The intermediate density 8-mers are organized into quintiles, where the their original bulk hydrogen configurations exhibit densities of 0.720, 0.876, 0.999, 1.14, and 1.25 g/cm$^3$, respectively. Results are presented in Figure~\ref{fig:mlip_extrapol}.

When the models are trained on both low $\rho$ and high $\rho$ 8-mers, errors below 0.1 eV/atom are also observed in all three MLIPs for the out-of-distribution intermediate density 8-mers. The lowest RMSEs are consistently observed for PET, which achieves average RMSEs of 0.0061 eV/atom for the energies and 0.148 eV/Å for the forces across all intermediate density quintiles.
Next best performance is observed for MACE, with an energy RMSE of 0.0228 eV/atom for the first quintile, below 0.02 eV/atom for all other quintiles, and an average force RMSE of 0.239 eV/Å across all quintiles. SOAP-BPNN exhibits average RMSEs of 0.0454 eV/atom and 0.674 eV/Å for the energies and forces, respectively, which are still well within an order of magnitude from the RMSEs observed for the original test set.

When the models are trained on the low $\rho$ 8-mers only, the out-of-distribution performance becomes worse for all three MLIPs. In fact, an overall increasing trend in the RMSEs is observed from the first intermediate density quintile (closest to low $\rho$) all the way to the high $\rho$ 8-mers. For MACE, RMSEs for the fifth quintile and high $\rho$ are \rev{much worse than} the other quintiles, exhibiting up to two orders of magnitude difference compared to the low $\rho$ 8-mers. For PET, the RMSEs are relatively more constant across the intermediate density quintiles and high $\rho$, with average values of 0.0296 eV/atom and 0.409 eV/Å for the energies and forces.
The analogous reverse trend is observed when the models are trained on the high $\rho$ 8-mers only, where the RMSEs for the intermediate $\rho$ and low $\rho$ 8-mers become significantly higher than before, and show an increasing trend from the fifth quintile (closest to high $\rho$) to the first quintile and low $\rho$.

Figure~\ref{fig:mlip_extrapol} also presents the RMSEs for the intermediate $\rho$ 8-mers when the body-order trends of the models are explicitly resolved to the DFT reference (Sec.~\ref{sec:bo_resolv_ref}). MACE exhibits performance degradation in all cases, and the degradation is far more pronounced when trained on the low or high $\rho$ 8-mers only. SOAP-BPNN shows reduced RMSEs for the case of high $\rho$ 8-mers only, and slight degradation in the other two cases. PET consistently displays slightly lower RMSEs for the out-of-distribution 8-mers when trained on the body-order-resolved datasets. This showcases the flexibility of PET to learn simultaneously the energetics of target 8-mers as well as their sub-clusters, then use the extra information from the sub-clusters to achieve further improvements in the RMSEs. \rev{In the Supplementary Material, we show that these trends are generally robust to changes in the hyperparameters of MACE and PET, except when nonlinear interaction blocks are used in MACE, which mitigates the previously observed performance degradation with body-order resolution to the DFT reference.}

Altogether, the generalizability of the MLIPs does not correlate clearly with any specific convergence trends in their body-orders. The explicit resolution of the body-orders also does not bring forth dramatic improvements in the extrapolative performance of the MLIPs. This hints at the absence of any practical benefit in enforcing the models to infer fast converging body-orders or directly learn the reference body-ordered energetics. If anything, such complexities can limit the learning capacity and add strain to the training exercise, potentially degrading both in- and out-of-distribution performance.

\section{Conclusion and outlook}

In this study, we have carefully analyzed the behavior of three different MLIPs in terms of their ``body-orderedness'', examining the body-ordered energy and force trends for hydrogen clusters extracted from bulk simulations at different densities, comparing between DFT and the MLIPs trained on a number of different dataset makeups.
In the reference DFT calculations, as seen in many other systems, we first observed that the MBE of energy shows a non-converging, oscillatory behavior for both ``molecular'' (low density) and ``atomic'' (high density) hydrogen clusters. The effect cannot simply be dismissed as an artifact of DFT, as the trend is also reproduced in state-of-the-art DMRG calculations.
Even though the oscillatory behavior can be explained in terms of the choice of the isolated atom energy as the baseline, higher body-order terms are large also for force-based metrics, which are insensitive to the choice of the baseline. An analysis of the electronic structure of the fragments points to the strong spin correlations as the origin of high body-order terms.
%We find its origin in the high degree of spin-spin correlations of the atomic clusters, as well as the choice of the isolated atom energy as the ``baseline'' of the body-order series. The persistence of high-order terms occurs also when considering the magnitude of the body-ordered forces, which is independent of the baseline.

When trained exclusively on 8-mers extracted from realistic bulk structures, the MLIPs all learn an effective MBE that is far from the reference---without any adverse effect on their in-domain accuracy. While the effective MBE is largely arbitrary for all models, MACE tends to prioritize the use of lower body-orders for a fast-converging trend. When the sub-clusters are incorporated into the training set to explicitly resolve the body-orders to the reference, MACE and PET quickly converge to the reference body-ordered energetics, but the relatively low descriptive power of SOAP-BPNN limits its accuracy on the $m$-mers and hence its ability to learn the reference MBE. 
We also observe that explicit body-order resolution degrades the accuracy on the full structures of interest for MACE, while it does not for PET. Contrary to what one might expect, the fast decay of the effective MBE does not translate into more robust extrapolative behavior. Explicitly resolving the body-orders does not improve the extrapolative performance for SOAP-BPNN, degrades it for MACE, and improves it slightly for PET.

While our experiments focus on one comparatively simple system, they suggest that there is little value in targeting explicitly the MBE, or in designing models that implicitly favor learning a fast-decaying effective body-ordered decomposition.
On the contrary, it appears that an unconstrained architecture such as PET, that does not build upon a hierarchical expansion of the neighbor density correlations but simply aims to achieve a highly expressive approximation of the target, demonstrates consistently the best performance, both for in- and out-of-distribution tasks.

The ``paradox'' of the MBE is resolved by recognizing that models trained on reasonably stable structures don't have to reproduce the MBE of the target. The large deviations between the true and empirical trends is a simple consequence of the fact that, for a model trained on those reasonable structures, the fragments that appear in the decomposition are highly distorted and amount to extrapolative predictions. 
We also speculate that the tendency of MACE to privilege a fast-decaying effective body-ordered energy decomposition may be a consequence of the ``contamination'' of high-order correlations with low-body-order components. \rev{We have shown in the Supplementary Material that the newly proposed nonlinear interaction blocks of MACE can alleviate this effect to an extent.~\cite{Batatia2025}} Given the strategy to ``purify'' the body-ordered components of the closely related atomic cluster expansion~\cite{Ho2024}, it could be interesting to investigate the behavior of a ``purified MACE'' architecture, to verify our hypothesis and observe if there are any consequences for accuracy and transferability. 
Overall, our observations suggest that, despite being an attractive approximation framework, and despite the strong mathematical connection to many widely used MLIP frameworks, the body-order decomposition is not especially useful as a guiding principle to design MLIPs.

\rev{
\section*{Supplementary Material}
A supplementary PDF document is provided, which contains further details of the \textit{ab initio} calculations, RMSE values, and results of additional experiments with variations in the dataset, baseline energy $E_1$, and ML model hyperparameters.
}
\section*{Acknowledgments}
The authors thank Alexander Goscinski, Arslan Mazitov, and Gabor Csányi for the useful discussions. 
SC, MD and MC acknowledge support from a SNSF grant (project ID 200020\_214879). FB and MC were supported by a project of the Platform for Advanced Scientific Computing (PASC). 
FG received funding from the ERC CoG grant FIAMMA (grant ID 101001890).
Calculations relied on computer time from the EPFL HPC platform SCITAS and from the CSCS project (ID lp26).

\section*{Data Availability}
The model training inputs, datasets and analysis scripts and notebooks 
are provided in a Materials Cloud~\cite{Talirz2020} repository under the following DOI: \texttt{10.24435/materialscloud:q7-da}. % The spin-adapted DMRG results used in this study are available at: \url{https://github.com/jiangtong1000/HydrogenCluster.git}, including selected geometries and corresponding spin-resolved total energies.

\appendix

\section{Linear body-ordered models}\label{app:lin_bo_models}

In linear and kernel-based MLIPs with the locality ansatz, the descriptors are often built to describe atom-centered body-order correlations. One example is Gaussian Approximation Potential (GAP)~\cite{Bartok2010} model utilizing SOAP descriptors,~\cite{Bartok2013} which is a 3-body descriptor. In some special cases (e.g. moment-tensor potential~\cite{Shapeev2016}, ACE~\cite{drau19prb}), $\mathbf{x}_i$ of Eq.~\eqref{eq:mlip_locale} can be expressed in terms of multiple body-ordered components:

\begin{equation}\label{eq:bofeats}
\mathbf{x}_i = \bigoplus_{\nu=1}^{\nu_{\text{max}}} \;\sum_{j_1,...,j_\nu \ne i} \varphi^\nu (\mathbf{r}_{ij_1}, \dots, \mathbf{r}_{ij_\nu})f_c(r_{ij_1}, \dots, r_{ij_\nu})
\end{equation}

\noindent Local environments are described at successively increasing body-orders (here expressed in terms of $\nu$) by functions $\varphi^\nu$ that describe the body-order $\nu+1$, and these descriptors are concatenated together to obtain $\mathbf{x}_i$. Since there exist separate ``blocks'' corresponding to the different body-orders, same separation can be applied to the trained weights of the model, which results in Eq.~\eqref{eq:mlip_linear}.

In such models, body-ordered descriptors of Eq.~\eqref{eq:bofeats} often contain self-interacting terms, i.e.~contributions where $j_1=j_2$. In ACE, allowing for such self-correlations, which is sometimes referred to as the ``density trick'', is what guarantees favorable scaling with the number of atoms in the environments. Chong et al.~\cite{Chong2025} have shown that the presence of such self-interactions leads to a model learning behavior where the apparent correspondence between the summands of Eqs.~\eqref{eq:mbe} and \eqref{eq:mlip_linear} cannot be captured by the model. More recently, Ho et al.~\cite{Ho2024} proposed a purification operator that removes the self-correlation contributions from the ACE descriptors, which allows the model to recover the above-mentioned correspondence in some specific cases (see Section 5.1 of Chong et al.~\cite{Chong2025}).

\section{Body-ordered energetics of 8-mer before and after fragmentation}\label{app:bo_frag}

To further verify that the slowly converging body-ordered energetics observed from the quantum chemical calculations are not stemming from DFT artifacts, namely the self-interaction error, we consider the ``fragmentation'' of an 8-mer into two 4-mer fragments and the changes in the body-ordered energetics thereof at different theory-levels. The 8-mer of interest is constructed by first arranging four hydrogen atoms into a square within the $xy$-plane with bond lengths of 1.4 Å, then replicating this configuration, rotating it by 45 degrees, and offsetting the replica in the $z$-direction by a distance that also results in similar nearest neighbor bond lengths. Taking this as the pre-fragmentation configuration, we then generate the post-fragmentation configuration by further increasing the offset distance between the two replicas by 50 Å.

\begin{figure}[h]
    \centering
    \includegraphics[width=\columnwidth]{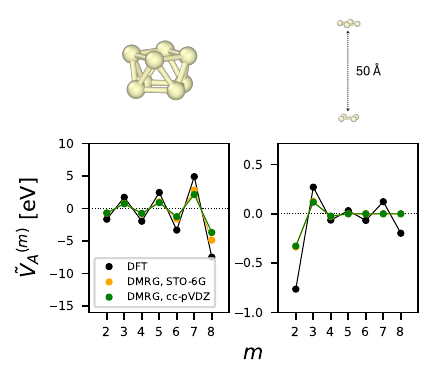}
    \caption{Body-ordered energetics of an 8-mer before (left) and after (right) fragmentation into two 4-mer fragments separated by 50 Å. Calculations are done with DFT as well as DMRG with STO-6G and cc-pVDZ basis sets. Corresponding atomic configurations are shown on top of the plots. Note that the $y$-axis of the two plots are presented in different scales.}
    \label{fig:bo_frag}
\end{figure}

In Figure~\ref{fig:bo_frag}, similar oscillatory trend is observed across all calculations for the 8-mer before fragmentation. When the cluster is separated into fragments, both DMRG calculations show a strict convergence of $\tilde{V}_A^{(m)}$ to 0 for $m>4$, which is the physically reasonable trend. On the contrary, DFT results contain unphysical, residual contributions for $m>4$ that originate from the self-interaction error. These results prove that the prevalent oscillatory body-ordered energetics captures the true, physical MBE trends at the atomic level for covalent systems, and not a consequence of the DFT self-interaction error.

\section{Dependence of body-order convergence on baseline energy}\label{app:convergence}

The alternating behavior of the body-ordered energies might appear strange, but can easily be explained as follows. 
Begin by noting that $V^{(m)}$ can be expressed as the cohesive energy  of the $m$-mer cluster (i.e. $E^{(m)} - m \cdot E_1$) minus the sub-cluster contributions. 
Assume that this term is proportional to $m$, $E^{(m)} - m \cdot E_1\approx m\epsilon $ (which is true in the asymptotic limit, but not necessarily for smaller clusters). Then, it is easy to see that:
\begin{equation}
\begin{split}
V^{(2)} = &E^{(2)} - 2 E_1 \approx 2\epsilon,\\
V^{(3)} = &E^{(3)} - 3 E_1 - \sum_{i<j} V^{(2)}(\textbf{r}_i, \textbf{r}_j) \approx 3\epsilon - 6\epsilon = -3\epsilon\\
V^{(4)} = &E^{(4)} - 4 E_1 - \sum_{i<j} V^{(2)}(\textbf{r}_i, \textbf{r}_j) \\ & - \sum_{i<j<k} V^{(3)}(\textbf{r}_i, \textbf{r}_j, \textbf{r}_k) \approx 4\epsilon - 12\epsilon + 12\epsilon = 4\epsilon \\
\end{split}
\end{equation}
which corresponds to the alternating, diverging trend. 
The fact that this trend continues can be proved by induction: 
\begin{equation}
V^{(m)}=E^{(m)}-mE_1-\sum_{k=2}^{m-1}V^{(k)} \frac{m!}{k!(m-k)!}
\end{equation}
assuming that up to $(m-1)$th body-order $V^{(k)}\approx (-1)^k k\epsilon$, one sees that the summation evaluates to $m(1-(-1)^m)\epsilon$, so that indeed $V^{(m)}=(-1)^m m\epsilon$.

These considerations justify the oscillatory behavior observed for the body-ordered energies, and indicate that adjusting $E_1$ to an effective value that zeroes out $\epsilon$ would facilitate the convergence of the expansion---consistent with the derivation in Ref.~\citenum{Thomas2022} that assumes the definition of a system-specific, effective energy reference.

\section{Over-representation of low body-orders in MACE}\label{app:mace_overrep}
Here, we discuss two ``channels'' that lead to the over-representation of the low body-order contributions in MACE. First, consider the initial features of MACE, which are defined in terms of ACE based on the density trick, as discussed in Section~\ref{sec:MBEandBOMLIPs}.
Since this implies the presence of self-interaction terms in the formalism~\cite{drau19prb,musi+21cr}, all higher body-order features also contain effectively lower body-order contributions, which makes the descriptors deviate away from the strictly canonical expression of the body orders. That is, in the summation of Eq.~\eqref{eq:bofeats}, there exist terms where $j_k=j_l$.
Recognizing their presence, we can re-express the descriptors by decomposing their contributions, as done in the following for the example of $\nu = 2$:
\begin{equation}
    \sum_{j_1,j_2\ne i} \varphi^2 (\mathbf{r}_{ij_1}, \mathbf{r}_{ij_2})f_c(r_{ij_1}, r_{ij_2}) = x^2_{i,\text{self}} + x^2_{i,\text{pure}}
\end{equation}
where the self-interacting terms are the one such defined from the same atom $j_1 = j_2$, namely
\begin{equation}
    x^2_{i,\text{self}} := \sum_j \varphi^2 (\mathbf{r}_{ij}, \mathbf{r}_{ij})f_c(r_{ij}, r_{ij})
\end{equation}
while $x^2_{i,\text{pure}}$ contains all the remaining pure $\nu=2$ terms, such that $j_1 \neq j_2$. Because the summation in $ x^2_{i,\text{self}}$ runs over only one indexes, the self interaction term is a contribution with an effective lower body-order.
Extending this argument to every increasing $\nu$, this proves that the basic ACE features, grounded in the density trick, implies that every $\nu$-term contains all the lower body-orders. As such, also every feature in MACE have contains an over-representation of lower orders. 
Moreover, as shown in Ref.~\cite{duss+22jcp}~(see Table 2), expansion terms containing self-interaction give rise to ill-conditioned representation. 
Therefore, one can infer that the terms for lower body order terms are of magnitude which is comparable with the one of the pure ones.

The second channel is provided by the update function, containing also a residual connection, used in the MACE architecture as defined in Ref.~\cite{bata+22nips}, which both contribute to the over-representation of the lower body-orders in the effective descriptors. The update function is defined as (from Ref.~\cite{bata+22nips}, Eq.~(12)):

\begin{equation}
    \mathbf h^{(t+1)}_{i,L} = U(\mathbf{m}^{(t)}_{i,L},\, \mathbf h^{(t)}_{i,L}) :=  \mathbf{U}^{(t)}_{kL} \cdot \mathbf{m}^{(t)}_{i,L} + \mathbf{W}^{(t)}_{z_i, k L}\cdot \mathbf h^{(t)}_{i,L},
\end{equation}
where $U$ is the update function, $\mathbf h^{(t)}_{i,L}$ are the features of the model after the $t$-th step of message passing, $ \mathbf{m}^{(t)}_{i,L}$ are the messages at the same steps, containing higher body order terms, and $ \mathbf{U}^{(t)}_{kL}$ and $\mathbf{W}^{(t)}_{z_i, k L}$ are learnable weights. At the core of the computation of the messages there are the same ACE contractions with self-interactions discussed above, and thus they have an over-representation of the lower body-orders.
Moreover, in defining the new features, the skip connections allow to utilize also the previous ones, which themselves contain the over-representation of lower body-orders.

% Finally, the readout procedure (see Ref.~\cite{bata+22nips}, Eq.~(13)) takes all the scalar features $\{\mathbf h^{(t+1)}_{i,0}\}$ for the prediction, which means that this over-representation is further amplified also at the readout level.

\bibliography{biblio.bib}

\end{document}

\title{Supplementary Material for \\ ``Resolving the Body-Order Paradox of Machine Learning Interatomic Potentials''}
\author{Sanggyu Chong}
\email{sanggyu.chong@epfl.ch}
\affiliation{Laboratory of Computational Science and Modeling, Institute of Materials, \'Ecole Polytechnique F\'ed\'erale de Lausanne, 1015 Lausanne, Switzerland}
\author{Tong Jiang}
\affiliation{Department of Chemistry and Chemical Biology, Harvard University, Cambridge, Massachusetts 02138, United States}
\author{Michelangelo Domina}
\affiliation{Laboratory of Computational Science and Modeling, Institute of Materials, \'Ecole Polytechnique F\'ed\'erale de Lausanne, 1015 Lausanne, Switzerland}
\author{Filippo Bigi}
\affiliation{Laboratory of Computational Science and Modeling, Institute of Materials, \'Ecole Polytechnique F\'ed\'erale de Lausanne, 1015 Lausanne, Switzerland}
\author{Federico Grasselli}
\affiliation{Laboratory of Computational Science and Modeling, Institute of Materials, \'Ecole Polytechnique F\'ed\'erale de Lausanne, 1015 Lausanne, Switzerland}
\affiliation{(Present address): Dipartimento di Scienze Fisiche, Informatiche e Matematiche, Università degli Studi
di Modena e Reggio Emilia, 41125 Modena, Italy}
\author{Joonho Lee}
\affiliation{Department of Chemistry and Chemical Biology, Harvard University, Cambridge, Massachusetts 02138, United States}
\author{Michele Ceriotti}
\affiliation{Laboratory of Computational Science and Modeling, Institute of Materials, \'Ecole Polytechnique F\'ed\'erale de Lausanne, 1015 Lausanne, Switzerland}

\newcommand{\FB}[1]{{\color{teal} #1}}
\newcommand{\FBcancel}[1]{{\color{teal} \st{#1}}}

\newcommand{\SC}[1]{{\color{orange} #1}}
\newcommand{\TJ}[1]{{\color{blue} #1}}
\newcommand{\MD}[1]{{\color{green} #1}}

\maketitle

\section*{Full details of cluster sampling protocol and first-principles calculations}

Among all bulk hydrogen configurations found in the Github repository of Cheng et al., configurations saved under the file name \texttt{`BQC-old-train-set-fps-n-5000.data'} were exclusively considered. This corresponds to 5000 configurations with 128 atoms selected via farthest point sampling performed on 200,000 snapshots of the \textit{ab initio} MD simulations. From these configurations, 100 with the highest density and 100 with the lowest density were taken for cluster sampling.

The cluster sampling algorithms were proposed to optimally capture the distinct chemistries of the two density regimes of bulk hydrogen. In the high density regime, bulk hydrogen is covalent and metallic, which results in significant correlations. As such, clusters were sampled by considering the nearest neighbors of a randomly selected central atom. In the low density regime, the system is insulating and mostly governed by intermolecular interactions between bonded dimers. To bets capture this, clusters were sampled by choosing a random atom and its nearest neighbor to complete the first bonded pair, then choosing additional \emph{pairs} of atoms that are the closest to the randomly sampled central atom.

The DFT calculations on the clusters and its sub-clusters were performed with FHI-aims (version 221103.1). PBE generalized gradient approximation exchange-correlation functional was used. The species default ``tight'' setting was used to define the basis set, integration grids, and the accuracy of the Hartree potential. Gaussian smearing with $\sigma=0.025$ eV was used to describe the occupations, where the value was chosen to smoothen out any potential discontinuities arising from significant differences in the ground state spin configurations along a smooth variation of the system coordinates, as probed in the dimer curve. The self-consistency threshold for the energy and density were set to be $10^{-6}$ eV and $10^{-6}$ e/\AA$^3$, respectively.

Density matrix renormalization group (DMRG) calculations are performed for selected hydrogen clusters to serve as benchmarks for understanding the true many-body behavior and to validate the accuracy of DFT calculations.
DMRG is a full configuration interaction solver that employs the low-rank 
matrix product state (MPS) approximation, which is well-suited for capturing strong correlation effects.
The method systematically truncates the Hilbert space while preserving the 
most important quantum entanglement, 
allowing for accurate treatment of systems that 
would be intractable with conventional quantum chemistry methods.
Notably, we tested coupled cluster with singles, doubles and triples (CCSDT), which is 
not as black-box as DMRG, since it relies heavily on the Hartree-Fock solution, which must be 
tested case by case, and even with a satisfactory HF solution, the converged CCSDT energy 
still does not converge to the FCI solution within 1 $mE_{\textrm{h}}$, highlighting the non-trivial nature of obtaining exact solutions when numerous configurations require careful manual attention and computational effort.
In our calculations, we perform DMRG with spin adaptation and enumerate 
the singlet, doublet, and triplet states for the subsystems involved in MBE, 
ensuring that we capture the correct ground state multiplicity for each cluster size.
The ground state energy is taken as the lowest one among these spin states,
which is crucial for systems where the ground state spin multiplicity may change
with cluster size or geometry.
The computations are performed with bond dimension $D$ up to 1000 to achieve 
convergence better than 0.1 $mE_{\textrm{h}}$. Both STO-6G and cc-pVDZ basis sets are investigated to assess basis set effects on the many-body expansion convergence and to establish the reliability of our
benchmark calculations across different levels of basis set completeness.
For STO-6G, we employ a single-particle
basis that is the set of Löwdin orthogonalized AO orbitals, and computed the spin-spin correlation among these localized orbitals.
For cc-pVDZ basis calculations, we use Hartree-Fock orbitals on the individual clusters.
\newpage

\renewcommand{\arraystretch}{1.5} 

\begin{table}[ht]
\centering
\caption{Test set RMSEs of MLIPs trained on both low $\rho$ and high $\rho$ hydrogen 8-mers.}
\begin{tabular}{C{3cm}|C{3cm}|C{3cm}|C{3cm}}
\hline
\textbf{RMSE} & SOAP-BPNN & MACE & PET \\
\hline
\hline
\textbf{Energy} (eV/atom) & 0.0242 & 0.0040 & 0.0024 \\
\hline
\textbf{Force} (eV/\AA) & 0.3672 & 0.0951 & 0.0665 \\
\hline
\end{tabular}
\end{table}

\begin{table}[ht]
\centering
\caption{Test set RMSEs of MLIPs trained on low $\rho$ hydrogen 8-mers only.}
\begin{tabular}{C{3cm}|C{3cm}|C{3cm}|C{3cm}}
\hline
\textbf{RMSE} & SOAP-BPNN & MACE & PET \\
\hline
\hline
\textbf{Energy} (eV/atom) & 0.0183 & 0.0018 & 0.0012 \\
\hline
\textbf{Force} (eV/\AA) & 2.3501 & 0.0601 & 0.0306 \\
\hline
\end{tabular}
\end{table}

\begin{table}[ht]
\centering
\caption{Test set RMSEs of MLIPs trained on high $\rho$ hydrogen 8-mers only.}
\begin{tabular}{C{3cm}|C{3cm}|C{3cm}|C{3cm}}
\hline
\textbf{RMSE} & SOAP-BPNN & MACE & PET \\
\hline
\hline
\textbf{Energy} (eV/atom) & 0.0232 & 0.0026 & 0.0033 \\
\hline
\textbf{Force} (eV/\AA) & 0.4055 & 0.0704 & 0.0719 \\
\hline
\end{tabular}
\end{table}

\newpage

\begin{figure}[h]
    \centering
    \includegraphics[width=\textwidth]{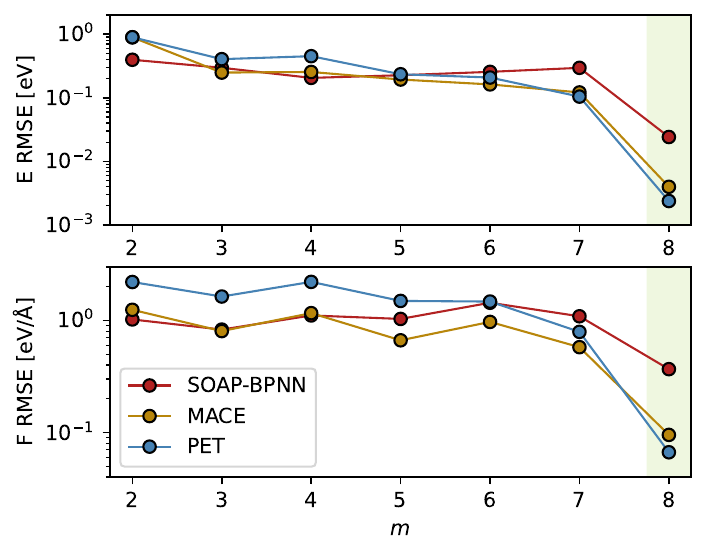}
    \caption{The RMSEs per $m$, where $2\le m \le 8$ for the three MLIPs trained on the datasets of low $\rho$ and high $\rho$ 8-mers, for 50 (25 low $\rho$ and 25 high $\rho$) of the test set 8-mers and their sub-clusters.}
    \label{suppfig:per-m-rmses}
\end{figure}

\newpage

\section*{Body-order trends with $E_1$ as a tunable parameter}

As discussed in Sections III and IV, the choice of the baseline energy $E_1$ can be varied from being fixed to the isolated atom energy as done in our main analysis. We note that such a choice, however, compromises the physical asymptotic behavior, which is to predict the (multiple of) isolated atom energies when each atomic environment in the system cannot perceive one another due to a separation larger than the radial cutoff. In SOAP-BPNN and MACE, a ``composition model'' is trained to handle the species-specific contributions to the potential energy surface, which is often obtained by training a linear regression model on the composition of the chemical system and the total reference energy. In our simple case of hydrogen clusters, this is equivalent to having $E_1$ as the average energy per atom across the training dataset. In PET, the self-attention mechanism allows the model to effectively fit to any value of $E_1$ during training.

In Figure S2, we observe that when the baseline energy E1 is freed up as a fitting parameter, SOAP-BPNN and MACE exhibit a relatively faster convergence trend in both the body-ordered energies and atomic forces. This is the expected behavior given the theoretical findings of Thomas et al. PET uniquely deviates from these trends, exhibiting relatively smaller energy contributions that persistently does not show a pattern of convergence, as well as relatively larger force contributions. These results are yet another manifestation of the fact that the PET architecture is not explicitly based on body-order correlations.

\begin{figure}[h]
    \centering
    \includegraphics[width=\columnwidth]{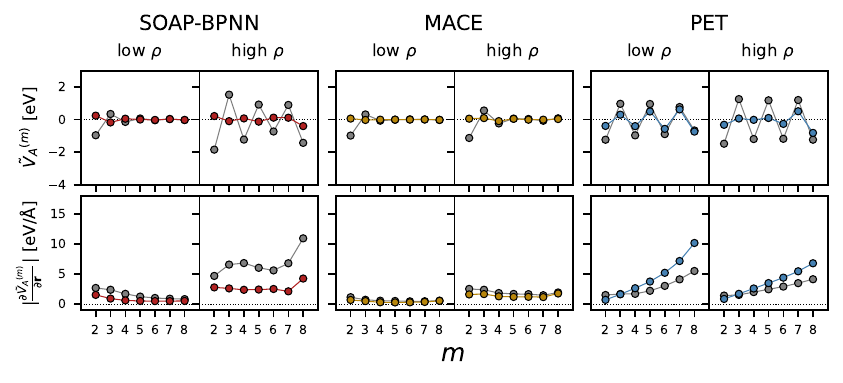}
    \caption{Body-order trends of the MLIPs trained with $E_1$ as a tunable parameter during model training. Mean $\tilde{V}^{(m)}_A$ and $| \partial \tilde{V}^{(m)}_A / \partial \textbf{r} |$ values over 25 low $\rho$ 8-mers and 25 high $\rho$ 8-mers are separately plotted. Gray markers correspond to the previous results with $E_1$ fixed to the isolated atom energy, and the colored markers correspond to the results when $E_1$ is freed during model training.}
    \label{suppfig:mlip_lc_rmse}
\end{figure}

\newpage

\begin{figure}[h]
    \centering
    \includegraphics[width=\columnwidth]{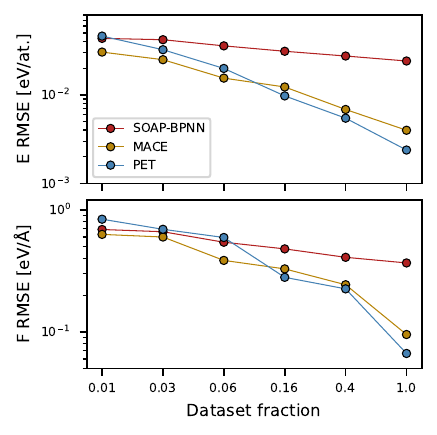}
    \caption{Learning curves of the MLIPs with respect to the dataset fraction, for the original dataset of both low and high $\rho$ 8-mers, presented in RMSEs for the energies (top) and the forces (bottom).}
    \label{suppfig:mlip_lc_rmse}
\end{figure}

\newpage

\begin{figure}[h]
    \centering
    \includegraphics[width=\textwidth]{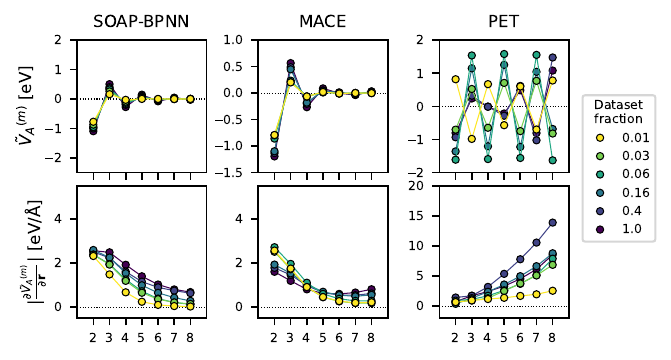}
    \caption{Body-order trends of the MLIPs trained on different fractions of a dataset with low $\rho$ 8-mers only. Mean $\tilde{V}^{(m)}_A$ and $| \partial \tilde{V}^{(m)}_A / \partial \textbf{r} |$ are plotted. Note that the $y$-axis ranges are tailored for each model to focus on their individual learning behavior.}
    \label{fig:mlip_lc_bo}
\end{figure}

\newpage

\begin{figure}[h]
    \centering
    \includegraphics[width=\columnwidth]{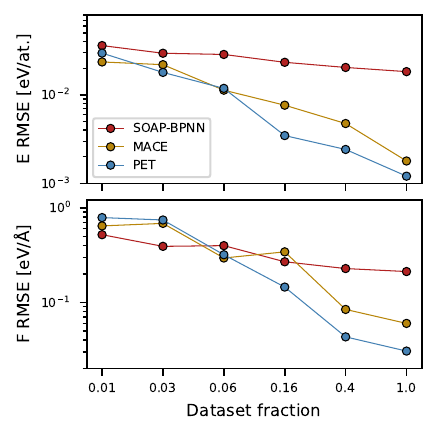}
    \caption{Learning curves of the MLIPs with respect to the dataset fraction, for the dataset of low $\rho$ 8-mers only, presented in RMSEs for the energies (top) and the forces (bottom).}
    \label{suppfig:mlip_lc_rmse}
\end{figure}

\newpage

\begin{figure}[h]
    \centering
    \includegraphics[width=\textwidth]{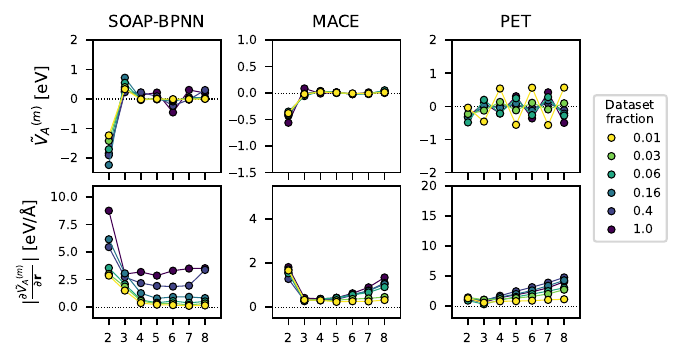}
    \caption{Body-order trends of the MLIPs trained on different fractions of a dataset with high $\rho$ 8-mers only. Mean $\tilde{V}^{(m)}_A$ and $| \partial \tilde{V}^{(m)}_A / \partial \textbf{r} |$ are plotted. Note that the $y$-axis ranges are tailored for each model to focus on their individual learning behavior.}
    \label{fig:mlip_lc_bo}
\end{figure}

\newpage

\begin{figure}[h]
    \centering
    \includegraphics[width=\columnwidth]{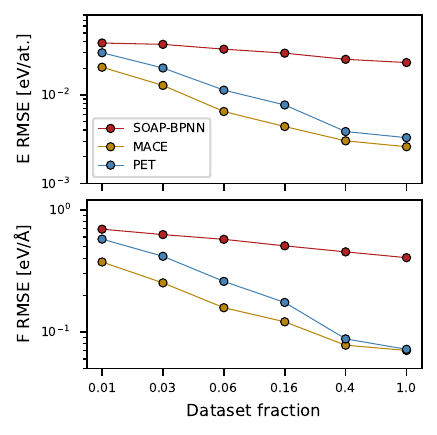}
    \caption{Learning curves of the MLIPs with respect to the dataset fraction, for the dataset of high $\rho$ 8-mers only, presented in RMSEs for the energies (top) and the forces (bottom).}
    \label{suppfig:mlip_lc_rmse}
\end{figure}

\newpage

\section*{Dependence of the body-order trends on the choice of model hyperparameters}

In this section, we present four case studies that further explore the effect of the model hyperparameters on the resulting body-order trends, for MACE and PET. In MACE, we also consider whether the changes in the hyperparameter choice improves its extrapolative performance especially when the body-orders are explicitly resolved within the dataset. Full set of model hyperparameters can be found in the model training inputs within the Materials Cloud repository associated with this work.

\subsection*{MACE: number of channels}
We tested the dependence of the body-order trends of MACE with respect to the number of channels tuned by the input parameter \texttt{num\_channels}. The original value was 128, and additional models were trained on the low and high density 8-mer dataset for the case where the number of channels is 64 and 256. Results in Figures S8 and S9 show that even when the number of channels is tuned from 128 to 64 or 256, no significant difference in the body-order trends is observed in both the body-ordered energies and forces, as well as the out-of-distribution performance on the middle density regime clusters.

\begin{figure}[h]
    \centering
    \includegraphics[width=\columnwidth]{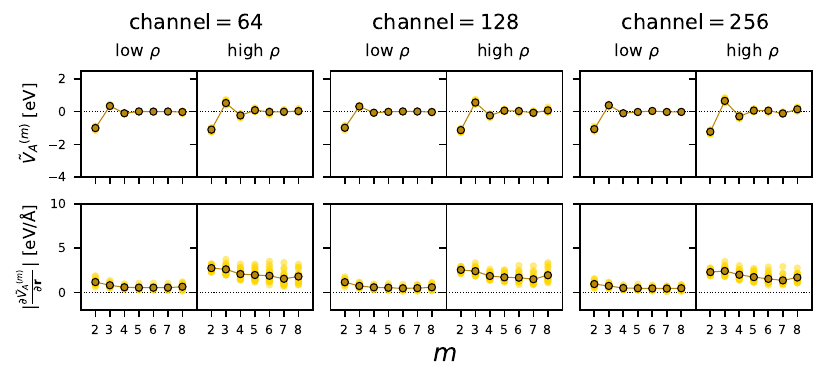}
    \caption{Average body-order trends of 25 low $\rho$ and 25 high $\rho$ hydrogen 8-mers computed with MACE, where the number of channels is varied from 128 to 64 and 256. The top row shows the results for energy ($\tilde{V}^{(m)}_A$) and the bottom row shows the results for the atomic forces ($| \partial \tilde{V}^{(m)}_A / \partial \textbf{r}|$).}
    \label{suppfig:mlip_lc_rmse}
\end{figure}

\newpage

\begin{figure}[h]
    \centering
    \includegraphics[width=\columnwidth]{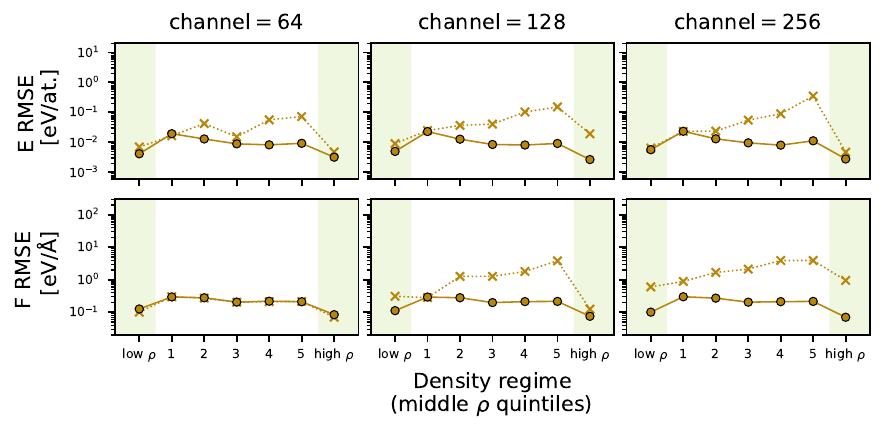}
    \caption{The energy (top) and force (bottom) RMSEs of MACE variants that differ in their number of channels, computed across the entire density range. $x$ axis denotes the different density regimes for which the RMSEs have been computed. The two extrema are the original low $\rho$ and high $\rho$ regimes used in model training, and the five enumerated middle ticks correspond to the quintiles of the out-of-distribution intermediate density regimes. Circles show the RMSEs for the models trained on the 8-mers only, and the crosses show the RMSEs for the models trained on the 8-mers and their sub-clusters.}
    \label{suppfig:mlip_lc_rmse}
\end{figure}

\newpage

\subsection*{MACE: correlation order $\nu$}
We tested the dependence of the body-order trends of MACE with respect to $\nu$, or the correlation order of the MACE internal features computed according to the ACE formalism. The original value was 3, and we trained new models on the low and high density 8-mer dataset with $\nu = 2, 4$. Results in Figures S10 and S11 reveal that $\nu$ controls the ``rate'' at which the body-order energy trend converges. In all MACE variants with different $\nu$ values, convergence in the body-ordered energies is consistently observed when $m > \nu + 1$. Despite the different convergence trends, no significant changes in the out-of-distribution performance is observed.

\begin{figure}[h!]
    \centering
    \includegraphics[width=\columnwidth]{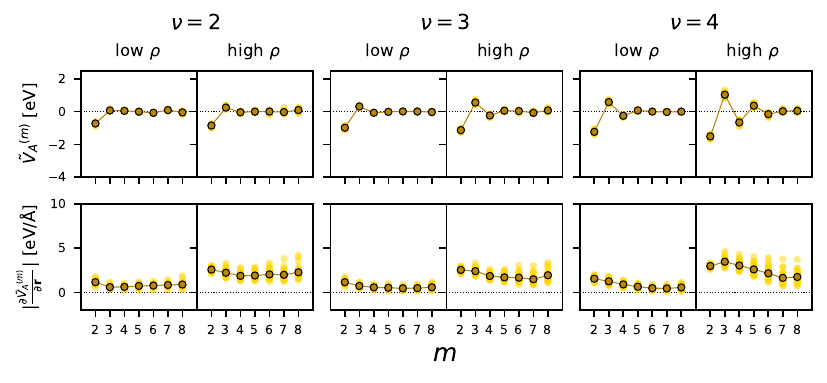}
    \caption{Average body-order trends of 25 low $\rho$ and 25 high $\rho$ hydrogen 8-mers computed with MACE, where the correlation order $\nu$ is varied from 3 to 2 and 4. The top row shows the results for energy ($\tilde{V}^{(m)}_A$) and the bottom row shows the results for the atomic forces ($| \partial \tilde{V}^{(m)}_A / \partial \textbf{r}|$).}
    \label{suppfig:mlip_lc_rmse}
\end{figure}

\newpage

\begin{figure}[h!]
    \centering
    \includegraphics[width=\columnwidth]{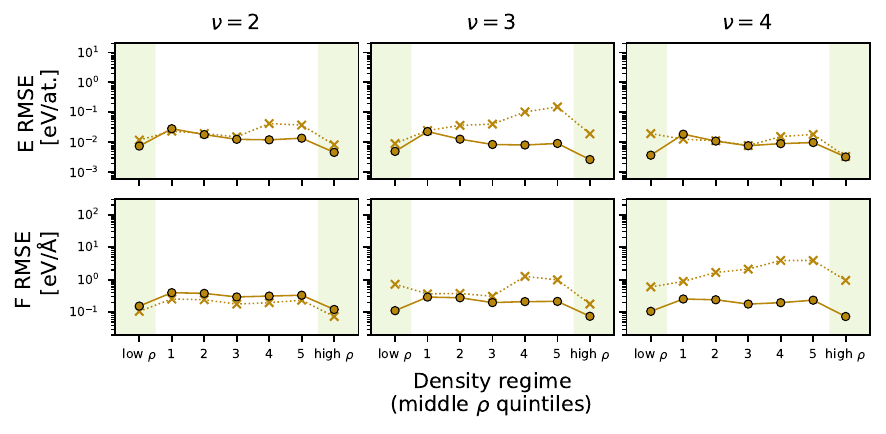}
    \caption{The energy (top) and force (bottom) RMSEs of MACE variants that differ in their correlation order $\nu$, computed across the entire density range. $x$ axis denotes the different density regimes for which the RMSEs have been computed. The two extrema are the original low $\rho$ and high $\rho$ regimes used in model training, and the five enumerated middle ticks correspond to the quintiles of the out-of-distribution intermediate density regimes. Circles show the RMSEs for the models trained on the 8-mers only, and the crosses show the RMSEs for the models trained on the 8-mers and their sub-clusters.}
    \label{suppfig:mlip_lc_rmse}
\end{figure}

\newpage

\subsection*{MACE: linear vs. nonlinear interaction blocks}
We tested the dependence of the body-order trends of MACE with respect to whether the linear or nonlinear interaction block is employed. The original MACE model employs linear interaction blocks, and we trained a new model on the low and high density 8-mer dataset with the nonlinear interaction blocks. The results in Figures S12 and S13 show that nonlinear interaction blocks allow the MACE architecture to relieve itself from the strictly converging body-order trends of the previous case. While no significant improvement or degradation is observed in the out-of-distribution performance, we do note that the nonlinear interaction blocks prevent the severe performance degradation with explicit resolution of the body-orders in the training dataset observed in the original model.

\begin{figure}[h!]
    \centering
    \includegraphics[width=0.75\columnwidth]{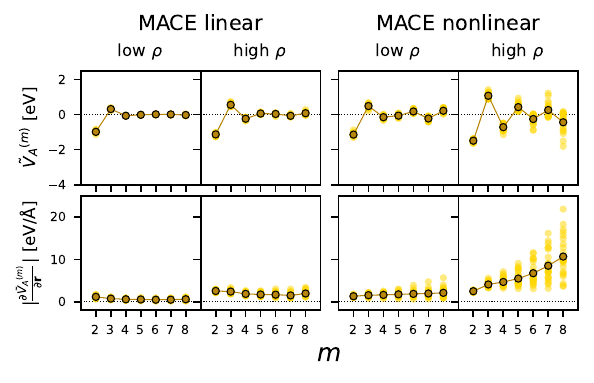}
    \caption{Average body-order trends of 25 low $\rho$ and 25 high $\rho$ hydrogen 8-mers computed with MACE variants that utilize either the linear interaction blocks (original, left) or the nonlinear interaction blocks (right). The top row shows the results for energy ($\tilde{V}^{(m)}_A$) and the bottom row shows the results for the atomic forces ($| \partial \tilde{V}^{(m)}_A / \partial \textbf{r}|$).}
    \label{suppfig:mlip_lc_rmse}
\end{figure}

\newpage

\begin{figure}[h!]
    \centering
    \includegraphics[width=0.75\columnwidth]{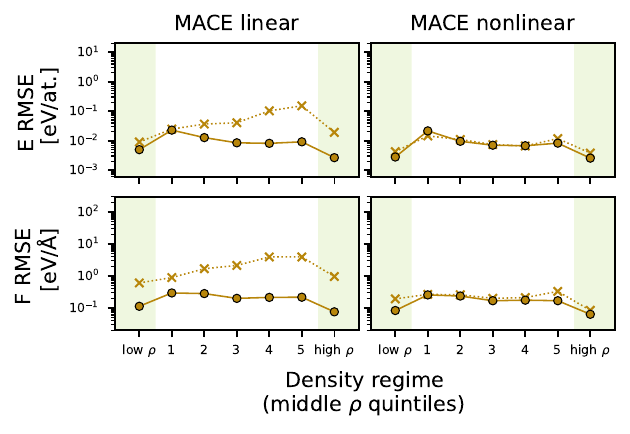}
    \caption{The energy (top) and force (bottom) RMSEs of MACE variants that differ in whether the use linear interaction blocks (original, left) or the nonlinear interaction blocks (right), computed across the entire density range. $x$ axis denotes the different density regimes for which the RMSEs have been computed. The two extrema are the original low $\rho$ and high $\rho$ regimes used in model training, and the five enumerated middle ticks correspond to the quintiles of the out-of-distribution intermediate density regimes. Circles show the RMSEs for the models trained on the 8-mers only, and the crosses show the RMSEs for the models trained on the 8-mers and their sub-clusters.}
    \label{suppfig:mlip_lc_rmse}
\end{figure}

\newpage

\subsection*{PET: $d_{\mathrm{PET}}$}

We tested the dependence of the body-order trends of PET with respect to `` $d_{\mathrm{PET}}$'', which is the dimension of the MLPs used for the PET edge interaction blocks. The original  $d_{\mathrm{PET}}$ value was 128, we trained new models on the low and high density 8-mer dataset with  $d_{\mathrm{PET}} = 64 \ \mathrm{and} \ 256$. The results in Figures S14 and S15 reveal that when  $d_{\mathrm{PET}}$ is altered, the model learns another arbitrary, non-converging body-order trend for the 8-mers. This is far more apparent for $d_{\mathrm{PET}} = 256$, where the trend becomes highly arbitrary and oscillatory, with the body-ordered energy and force contributions several orders of magnitude larger than the other two cases. This is explained by the fact that the number of parameters of the model increases from approximately 1 million to 2.8 million, which allows the model to achieve accurate fit to the presented 8-mer dataset (8-mer test set E RMSE is 0.0024 eV and F RMSE is 0.060 eV/\AA, comparable to the original results in Table S1) despite the severe arbitrariness in the inferred body-order decomposition trends. We note that the out-of-distribution performance of PET is not significantly affected by the changes in $d_{\mathrm{PET}}$.

\begin{figure}[h!]
    \centering
    \includegraphics[width=\columnwidth]{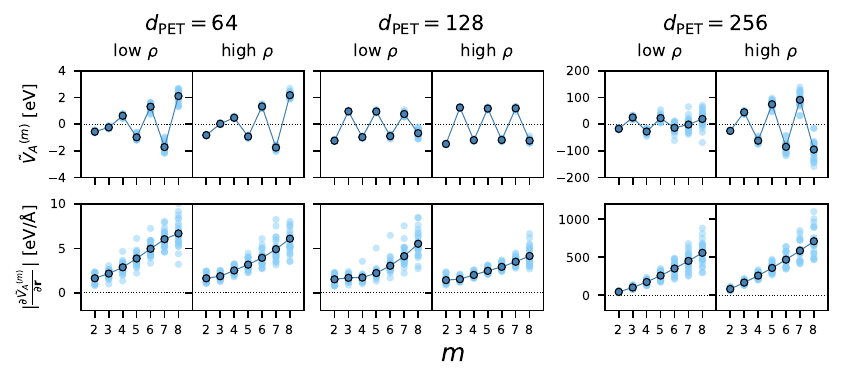}
    \caption{Average body-order trends of 25 low $\rho$ and 25 high $\rho$ hydrogen 8-mers computed for PET trained with different values of $d_{\mathrm{PET}}$. The original PET model used $d_{\mathrm{PET}} = 128$, and new results are shown for $d_{\mathrm{PET}} = 64, 256$. The top row shows the results for energy ($\tilde{V}^{(m)}_A$) and the bottom row shows the results for the atomic forces ($| \partial \tilde{V}^{(m)}_A / \partial \textbf{r}|$).}
    \label{suppfig:mlip_lc_rmse}
\end{figure}

\newpage

\begin{figure}[h!]
    \centering
    \includegraphics[width=\columnwidth]{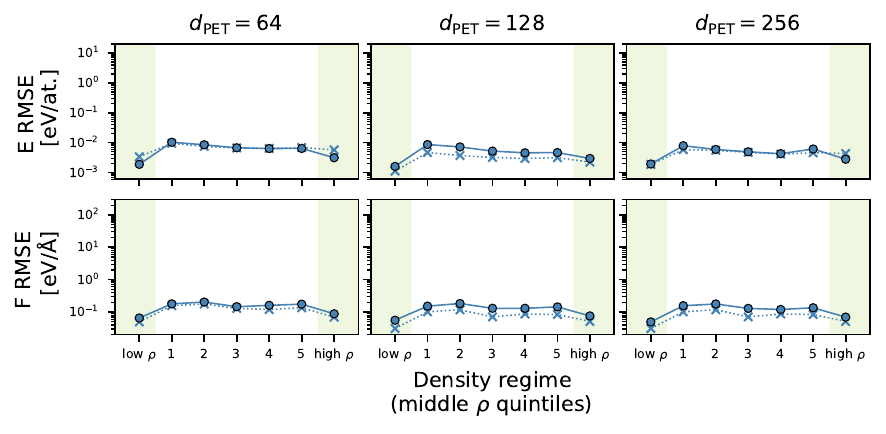}
    \caption{The energy (top) and force (bottom) RMSEs of PET variants that differ in their $d_{\mathrm{PET}}$ hyperparameter. Original model used $d_{\mathrm{PET}} = 128$, and the variants have been trained with $d_{\mathrm{PET}} = 64, 256$. $x$ axis denotes the different density regimes for which the RMSEs have been computed. The two extrema are the original low $\rho$ and high $\rho$ regimes used in model training, and the five enumerated middle ticks correspond to the quintiles of the out-of-distribution intermediate density regimes. Circles show the RMSEs for the models trained on the 8-mers only, and the crosses show the RMSEs for the models trained on the 8-mers and their sub-clusters.}
    \label{suppfig:mlip_lc_rmse}
\end{figure}